\documentclass[aps,showpacs,dvips,showkeys,preprintnumbers,nofootinbib]{revtex4}
\usepackage{graphicx}
\usepackage{color}
\usepackage{amsmath}
\usepackage{amssymb}

\def \be  {\begin{equation}}
\def \ee  {\end{equation}}
\def \ee  {\end{equation}}
\def \bea {\begin{eqnarray}}
\def \eea {\end{eqnarray}}

\def \Tr  {\bf{Tr}}

\begin{document}

\preprint{ECTP-2018-02}    
\preprint{WLCAPP-2018-02}
\hspace{0.05cm}
\title{Transverse Momentum Spectra of Pions at LHC Energies}

\author{Abdel Nasser Tawfik}
\email{atawfik@nu.edu.eg}
\affiliation{Nile University - Egyptian Center for Theoretical Physics, Juhayna Square of 26th-July-Corridor, 12588 Giza, Egypt}
\affiliation{World Laboratory for Cosmology And Particle Physics (WLCAPP), 11571 Cairo, Egypt}

\date{\today}

\begin{abstract} 

In order to characterize the transverse momentum spectra of positive pions measured in the ALICE experiment, two thermal approaches are utilized; one is based on degeneracy of non-perfect Bose-Einstein gas and the other imposes an {\it ad-hoc} finite pion-chemical potential. The inclusion of missing haron states and the out-of-chemical equilibrium greatly contribute to the excellent characterization of pion production. The excellent reproduction of the experimental data can be understood as a manifestation of not-yet-regarded anomalous pion-production, which likely contribute to the long-standing debate on the {\it ''anomalous''} proton-to-pion ratios at top RHIC and LHC energies.

\end{abstract}

\pacs{13.85.Ni, 05.70.Ln, 13.87.Fh}
\keywords{Inclusive production with identified hadrons, Nonequilibrium and irreversible thermodynamics, Fragmentation into hadrons}

\maketitle

\section{Introduction}

The collective properties of strongly interacting matter (radial flow, for instance) and dynamics of colliding hadrons can be explored from the study of transverse momentum distributions $p_T$ of produced particles. RHIC results on well-identified particles produced at low $p_T$, especially pions, have shown that the bulk matter created can be well described by hydrodynamics \cite{McLerran2004}. It should be emphasized that the high-$p_T$ spectra, especially for the lowest-lying Nambu-Goldstone bosons, the pions, likely manifest dynamics and interactions of partons and jets created in the earliest stage of nuclear collisions \cite{BauerWestfall}. For instance, the collective expansion in form of radial flow might be caused by internal pressure gradients. Furthermore, the $p_T$ distributions are assumed to determine conditions, like temperature and flow velocity, gaining dominance during the late evolution of the collision which is generically described as kinetic freeze-out, where the elastic interactions are ceased, conclusively. 

From the theoretical point of view, the $p_T$-spectra are excellent measurements enabling us a better understanding of the QCD interactions. Soft nonperturbative QCD can be well applied to low $p_T$ regime (below a few GeV/c) \cite{Wong2013}. Fragmentation of QCD string \cite{Andersson87}, parton wave functions in flux tube \cite{Wong94}, parton thermodynamics \cite{Kraus2009} and parton recombination \cite{Wibig2010} are examples on underlying physics. At high-$p_T$, hard-scattering cross section from QCD perturbative calculations, parton distribution functions, and parton-to-hadron fragmentation functions have been successfully utilized in reproducing $p_T$-spectra of various produced particles \cite{Brodsky2012}.

It is worth mentioning that there is no a well-defined line separating nonperturbative from perturbative $p_T$-regimes \cite{Wong2013}. Various theoretical studies are not even able to distinguish between both of them. For instance, when constructing partition functions, extensive and nonextensive statistical approaches are frequently mistaken \cite{Tawfik:2018ahq,Tawfik:2017bsy,Tawfik:2016pwz}. The claim that high $p_T$-spectra of different produced particles are to be reproduced by Tsallis statistics seems incomplete \cite{Tawfik:2016pwz,Bialas2015}. This simply inspires a great contradiction between nonperturbative and perturbative QCD \cite{Bialas2015}. The statistical cluster decay could be scaled as power laws very similar to the ones of Tsallis statistics. The earlier is conjectured to cover a wide range of $p_T$, while the later is limited to a certain $p_T$ regime. This would lead to an undesired mixing up that the observed power-laws might be stemming from the statistical cluster decay and interpreted due to Tsallis-type of nonextensivity.

In additional to the proposal of utilizing a generic (non)extensive statistical approach \cite{Tawfik:2018ahq,Tawfik:2017bsy,Tawfik:2016pwz}, we want here to recall another theoretical framework based on an {\it ad-hoc} physically motivated assumption that the pion production might be interpreted due an out-of-equilibrium process \cite{ruuskanen90}. Such an approach is stemming from the pioneering works of Bogolubov devoted to an explanation for the phenomenon of superfluidity on the basis of degeneracy of a non-perfect Bose-Einstein gas \cite{Bogolubov1947} and determining the general form of the energy spectrum; an ingenious application of the second quantization \cite{Landau1949}. 

After a short review of the thermal approach, the Hadron Resonance Gas (HRG) model in equilibrium and discussing on how to drive it towards non-equilibrium through inclusion of repulsive interactions, Sec. \ref{sec:hrgEqlbm}. In Sec. \ref{sec:NoNhrgEqlbm}, we elaborate modifications carried out towards implementing non-perfect Bose-Einstein gas based on a proposal of pion superfluidity. Another out-of-chemical-equilibrium thermal approach shall be outlined in Sec. \ref{sec:thrmlNoNhrgEqlbm}, where finite pion-chemical potential is {\it ad hoc} imposed. The results shall be discussed in Sec. \ref{sec:Rslts}. The conclusions shall be given in Sec. \ref{sec:Cncl}.

\section{A short Review on Equilibrium Resonance Gas with Van der Waals}
\label{sec:hrgEqlbm}

The hadron resonances treated as a non-interacting gas \cite{Karsch:2003vd,Karsch:2003zq,Redlich:2004gp,Tawfik:2004sw,Tawfik:2004vv} are conjectured to determine the equilibrium thermodynamic pressure of QCD matter below chiral and deconfinement critical temperature, i.e. hadronic phase. It has been shown that the thermodynamics of a strongly interacting system can also be approximated as an ideal gas composed of hadron resonances with masses $\lesssim 2~$GeV \cite{Tawfik:2004sw,Vunog}. Interested readers are kindly advised to consult the most recent review article \cite{Tawfik:2014eba}. 

The grand canonical partition function can be constructed as
\bea
Z(T, \mu, V) &=& \Tr \left[ \exp^{\frac{\mu\, N-H}{T}}\right],
\eea
where $H$, $T$, and $\mu$ are the Hamiltonian, the temperature, and the chemical potential of the system, respectively. The Hamiltonian can be given by as summation of the kinetic energies of relativistic Fermi and Bose particles including the relevant degrees-of-freedom and the interactions resulting in formation of resonances and well describing the particle production in high-energy hadron collisions. Under these assumptions, the sum over the {\it single-particle partition} functions $Z_h^1$ of existing hadrons and their resonances introduces dynamics to the partition function 
\bea 
\ln Z(T, \mu_h ,V) &=& V \sum_h\pm \frac{g_h}{2\pi^2}\int_0^{\infty} k^2 dk \ln\left\{1 \pm \exp\left[\frac{\mu_h -\varepsilon_h}{T}\right]\right\}, \label{eq:lnz1}
\eea
where $\varepsilon_h=(k^2+ m_h^2)^{1/2}$ is the $h-$th particle dispersion relation, $g_h$ is
spin-isospin degeneracy factor and $\pm$ stands for fermions and bosons, respectively.

In the present work, we include only the known hadron resonances with masses $\leq 2~$GeV compiling by the particle data group (PDG) 2018 \cite{PDG2018}. This mass cut-off is assumed to define the validity of the HRG model in characterizing the hadron phase \cite{hgdrn}. The inclusion of hadron resonances with heavier masses leads to divergences in all thermodynamic quantities  expected at temperatures larger than the Hagedorn temperature~\cite{Karsch:2003zq,Karsch:2003vd}. In addition to these aspects, there are fundamental reasons (will be elaborated in forthcoming sections) favoring the utilization of even {\it ideal} HRG model in predicting the hadron abundances and their thermodynamics. For the seek of completeness, we highlight that the hadronic resonances which not yet measured, a parametrization for total spectral weight has been introduced \cite{brnt}. 

As given earlier, we assume that the constituents of the HRG are free (collisionless) particles. Some authors prefer taking into account the repulsive ({\it electromagnetic}) van der Waals interactions in order to partly compensate strong interactions in the hadronic medium \cite{Tawfik:2013eua} and/or to drift the system towards even partial non-equilibrium. Accordingly, each constituent is allowed to have an {\it eigen}volume and the hadronic system of interest becomes thermodynamically partially out-of-equilibrium\footnote{How does a statistical thermal system, like HRG, become out-of-equilibrium? To answer this question, one might need to recall the main parameters describing particle production in equilibrium. These are $T$, $\mu_h$ and $V$ \cite{Floris}. In the present script, we first focus on the third parameter, and therefore describe this as a partial out-of-equilibrium process. The volume $V$, the normalization parameter typically constrained by pions, becomes a subject of modification through van der Waals repulsive interactions, for instance. Furthermore, it should be also noticed that the chemical potential $\mu_h$ should be modified, as well, at least in connection with the modification in $V$. This would explain that taking into account van der Waals repulsive interactions contributes to deriving the system of interest towards non-equilibrium.}. Thus, the total volume of HRG constituents should be subtracted from the fireball volume or that of heat bath. Considerable modifications in thermodynamics of HRG including energy, entropy, and number densities should be taken into consideration. The hard-core radius of hadron nuclei can be related to the multiplicity fluctuations. 

How large modified $V$ is allowed to be? This would be determined for the capability of the HRG model with finite-volume constituents to reproduce relaible lattice QCD thermodynamics \cite{Tawfik:2013eua}. At radius $r>0.2~$fm, the disagreement with the first-principle QCD calculations becomes more and more obvious and increases with the increase in the radii. It was concluded that such an excluded volume-correction becomes practically irrelevant, as it comes up with negligible effects at low temperatures  \cite{Tawfik:2013eua}. But on the other hand, it shows a remarkable deviation from the lattice QCD calculations, especially at high $T$. 

The repulsive interactions between hadrons are considered as a phenomenological extension and exclusively based on van der Waals excluded volume \cite{exclV1,exclV2,exclV3,exclV4}.  Intensive theoretical works have been devoted to the estimation of the excluded volume and its effects on the particle production and fluctuations \cite{exclV5}, for instance. It is conjectured that the hard-core radius of hadron nuclei can be related to the multiplicity fluctuations \cite{exclV6}. In the present work, we simply assume that hadrons are spheres and all have the same radius. On the other hand, the assumption that the radii would depend on the hadron masses and sizes could come up with a very small improvement.

Although various types of interactions have been assumed, as well \cite{Stoecker2018,Stoecker2018B}, we focus on the van der Waals repulsive interaction. By replacing the system volume $V$ by the actual one $V_{act}$, the van der Waals excluded volume can be deduced \cite{exclV1}
\begin{eqnarray}
V_{act} &=& V - \sum_h\, v_h\, N_h, \label{eq:evc1}
\end{eqnarray}
where the volume and particle number of each constituent hadron are $v_h=4\, (4 \pi r_h^3/3)$ and $N_h$, respectively. $r_h$ is the corresponding hard sphere radius of $h$-th particle. The procedure encoded in Eq. (\ref{eq:evc1}) leads to modification in the chemical potential $\tilde{\mu}_h=\mu_h - v_h\, p$, where the thermodynamic pressure $p$ is self-consistently given as $\sum_h p_h^{id}(T,\tilde{\mu}_h)$ and 
\begin{eqnarray}
n &=& \frac{\sum_h n_h^{id}(T,\tilde{\mu}_h)}{1+\sum_h v_h n_h^{id}(T,\tilde{\mu}_h)}, \\
\epsilon &=& \frac{\sum_h \epsilon_h^{id}(T,\tilde{\mu}_h)}{1+\sum_h v_h n_h^{id}(T,\tilde{\mu}_h)}, \\
s &=& \frac{\sum_h s_h^{id}(T,\tilde{\mu}_h)}{1+\sum_h v_h n_h^{id}(T,\tilde{\mu}_h)},
\end{eqnarray}
where the superscript $id$ refers to thermodynamic quantities calculated in HRG model with point-like constituents.
  
In the section that follows, we discuss on out-of-chemical-equilibrium $p_T$ spectra of positive pions, where finite pion chemical potential should be {\it ad hoc} inserted in.

\section{Out-of-Equilibrium $p_T$-Spectra of Pions}  
\label{sec:NoNhrgEqlbm}

In U($1$) global symmetry, where the scalar field $\phi(x)$ has a unitary transformation by the phase factor $\exp(-i \alpha)$, the Bose-Einstein condensation of lowest-lying Nambu-Goldstone bosons could be studied from the partition function \cite{kapusta}
\bea
\ln z(T,\mu_{\pi}) &=& \frac{V}{T} \left(\mu_{\pi}^2-m^2\right) \xi^2 - V \int \frac{d^3 p}{(2\, \pi)^3} \left[\frac{\varepsilon}{T} + \ln\left(1-e^{-\frac{\varepsilon-\mu_{\pi}}{T}}\right) + \ln\left(1-e^{-\frac{\varepsilon+\mu_{\pi}}{T}} \right)\right],
\eea
where $\xi$ is a parameter carrying full infrared characters of the scalar field. This can be treated as a variational parameter relating to the charge of condensed boson particle. At $|\mu_{\pi}|<m$, Eq. (\ref{eq:lnz1}) can obviously be recovered. At $\mu_{\pi}\rightarrow m$, the transverse momentum spectrum of pions is given as
\bea
\frac{1}{2 \pi\, p_T} \frac{d^2\, N_{\pi}}{d p_y\, d y} &=& V \frac{m_T^{(\pi)}\, \cosh(y)}{(2 \pi)^3} \left\{\exp\left[\frac{m_T^{(\pi)}\, \cosh(y) - \mu_{\pi}}{T}\right]-1\right\}^{-1}\nonumber \\
&+&  V \frac{m_T^{(reson.)}\, \cosh(y)}{(2 \pi)^3} \left\{\exp\left[\frac{m_T^{(reson.)}\, \cosh(y) - \mu_{\pi}}{T}\right]-1\right\}^{-1}\, \times \mathtt{Branching}\; \mathtt{Ratio}^{(reson.)}, \label{eq:pionpTa}
\eea
where $m_T=(p_T^2+m^2)^{1/2}$ is the pion transverse mass. The volume element $d^3 p$ was expressed in ($p_T$), rapidity ($y$) and azimuthal angle ($\phi$) as $d^3 p=p_T m_T \cosh(y) d p_T\, d y\, d \phi$ and the energy becomes $\varepsilon=m_T\, \cosh(y)$.

The pion $p_T$-spectrum is calculated from the single-particle expression (\ref{eq:pionpTa}) plus all contributions coming from the heavier hadron resonances which decay into pions, in which the corresponding branching ratio should be taken into consideration. In the results shown in Fig. \ref{fig:ptSpect1}, we distinguish between the hadron resonances with the given mass cut-off and that without $\sigma$-states. In both cases, we also distinguish between results at chemical equilibrium of pion production, i.e. $\mu_{\pi}=0$, and that at out-of-equilibrium, i.e. $\mu_{\pi}\neq0$.

For hadron masses $<2~$GeV, various sigma states are in(ex)cluded: $I=1/2$, $K_0^*(800)$ known as $\kappa$, which was excluded particle data group 2014 and from our calculations as well and $K_0^*(1430)$, $I=1$, $a_0(980)$ and $a_0(1450)$ and $I=0$, $f_0(500)$ widely known as $\sigma$, $f_0(980)$, $f_0(1370)$, $f_0(1500)$, and $f_0(1710)$. It intends to investigate the importance of sigma states in reproducing $p_T$-spectra of pions at LHC energies.

\section{Out-of-Equilibrium Thermal Distribution}
\label{sec:thrmlNoNhrgEqlbm}

In this section, we recall another thermal distribution at out-of-chemical-equilibrium \cite{ruuskanen90}, inspired by the proposal of Bogolubov to explain the phenomenon of superfluidity as degeneracy of a non-perfect Bose-Einstein gas and determine a general form of energy spectra; a kind of ingenious application of second quantization. 

In the approach of ref. \cite{ruuskanen90}, the $p_T$-spectra of {\it negatively} charged bosons from $200~$A~GeV O+Au and S+S collisions by NA35 experiment have been successfully fitted. It was assumed that a cylindrical tube of matter with radius $R$ expands, longitudinally, but without transverse flow $\nu_z=z/t$. The $p_T$-distribution could be determined, when replacing $\varepsilon$ in Eq. (\ref{eq:lnz1}), for instance, by the azimuthal angle $\phi$ and the covariant form $p_{\mu}\, u^{\mu}$ (four-momentum and -velocity) and then integrating over the freeze-out time $\tau_{fo}=\tau$. 

Then, at finite rapidity $y\neq0$, the pion $p_T$-spectrum reads
\bea
\frac{1}{2 \pi\, p_T} \frac{d^2\, N_{\pi}}{d p_y\, d y} &=& \left(\pi\, R^2\, \tau_{fo}\right) \frac{m_T\, \cosh(y)}{(2 \pi)^3} \sum_{n=1}^{\infty} (\pm)^{n+1}\, \exp\left(n\frac{\mu_{\pi}}{T}\right)\, K_1\left[n\frac{m_T}{T} \cosh(y)\right]. \label{eq:pionpTb}
\eea
The contributions likely added by heavy resonances can straightforwardly be contributing to Eq. (\ref{eq:pionpTb}). In Fig. \ref{fig:ptSpect1}, we distinguish between results at chemical equilibrium of pion production, i.e. $\mu_{\pi}=0$ and that at out-of-chemical equilibrium, i.e. $\mu_{\pi}\neq0$.

\section{Results}
\label{sec:Rslts}

\begin{figure}[hbt]
\includegraphics[width=3.5cm,angle=-90]{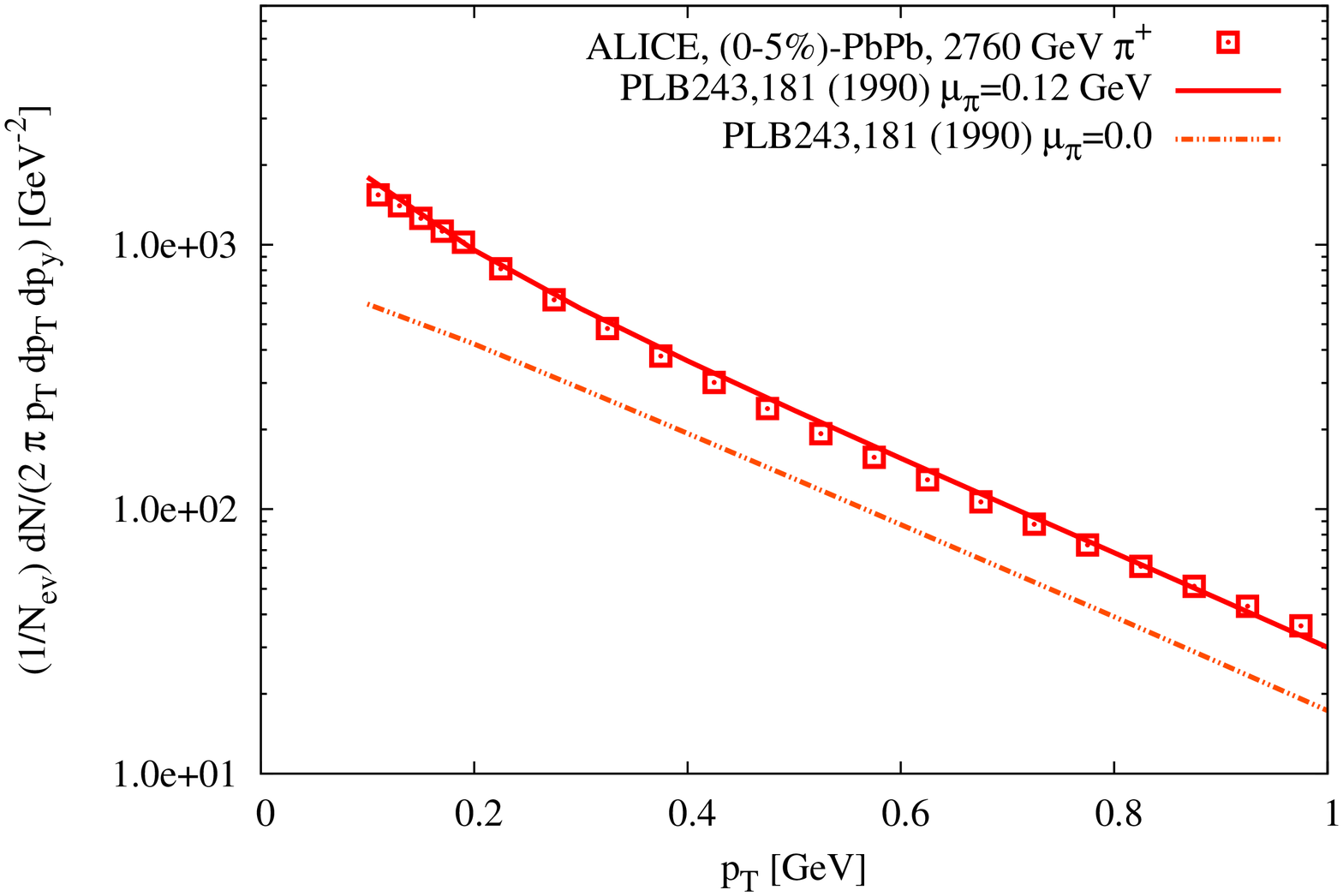}
\includegraphics[width=3.5cm,angle=-90]{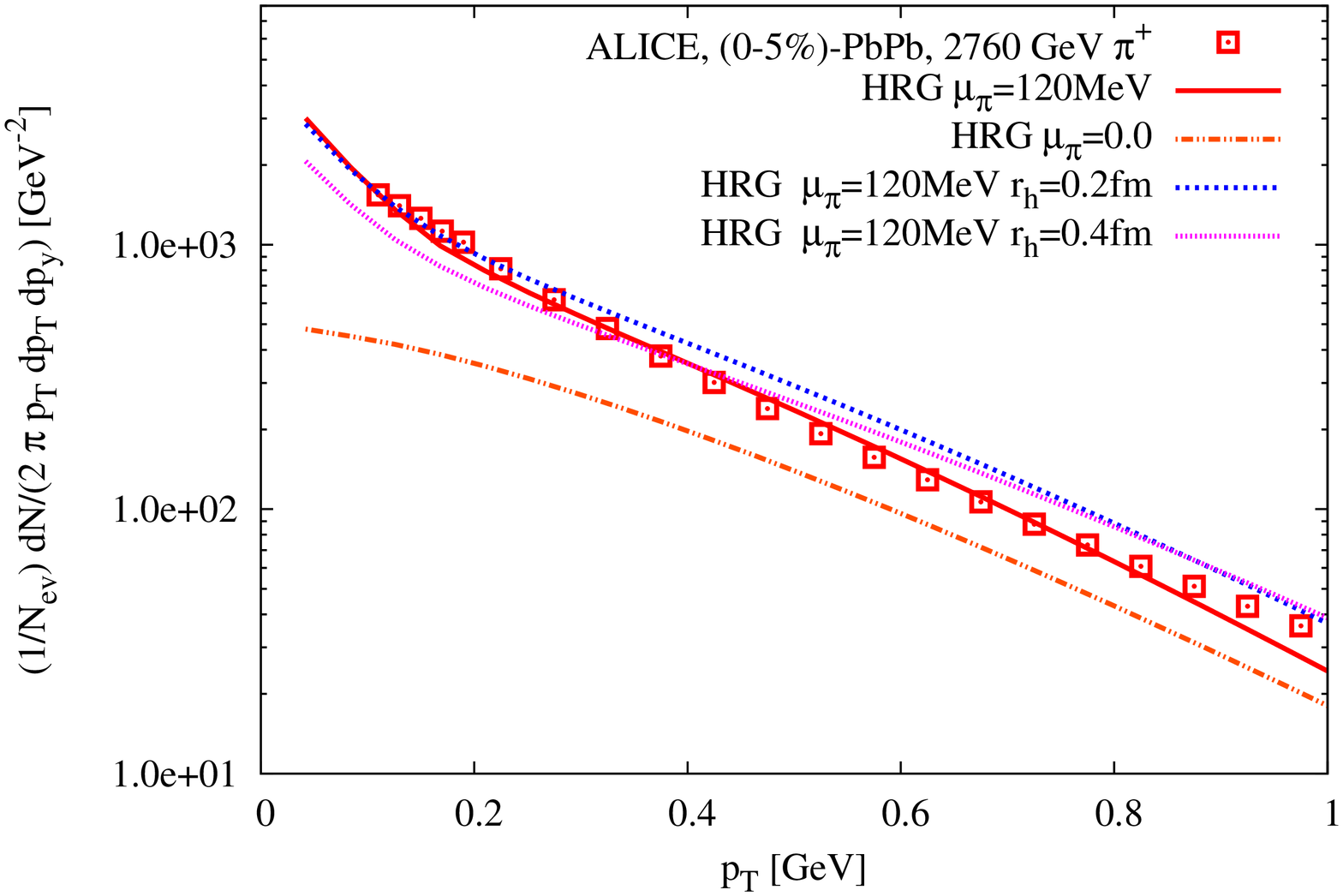}
\includegraphics[width=3.5cm,angle=-90]{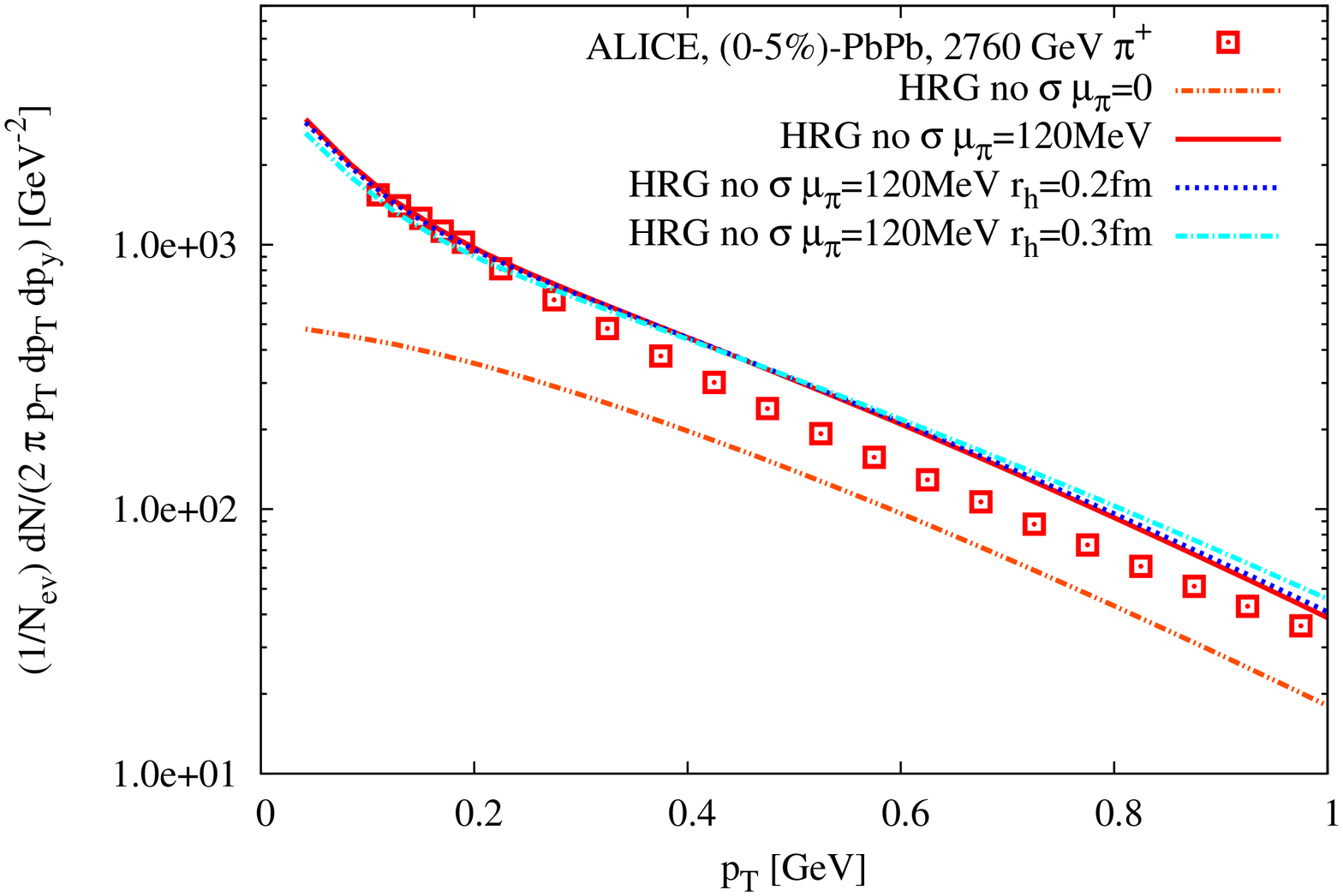}
\caption{Number of positive pions per transverse momentum per rapidity is given dependence on the transverse momentum. The symbols refer to the ALICE measurements in the most central ($0-5\%$) Pb+Pb collisions at $2.67~$TeV. Left-hand panel shows a comparison with Eq. (\ref{eq:pionpTb}) at vanishing $\mu_{\pi}=0$ and finite $\mu_{\pi}=0.12~$GeV \cite{AliceSpectra}. In middle and right-hand panels, HRG calculations, Eq. (\ref{eq:pionpTb}), with and without scalar $\sigma$ states and excluded-volume corrections are confronted to the experimental measurements, as well. \label{fig:ptSpect1} }
\end{figure}

Figure \ref{fig:ptSpect1} shows $p_T$-spectra of positive pions in dependence on the transverse momentum. The symbols are the ALICE measurements in the most central ($0-1\%$) Pb+Pb collisions at $2.67~$TeV \cite{AliceSpectra}. The calculations from Eq. (\ref{eq:pionpTb}) at vanishing (dash-double-dotted curve) and finite $\mu_{\pi}=0.12~$GeV (solid curve), left panel. The HRG calculations, Eq. (\ref{eq:pionpTa}) with and without scalar sigma states are illustrated in middle and right panels, respectively. Also, here we distinguish between vanishing (dash-double-dotted curve) and finite $\mu_{\pi}=0.12~$GeV (solid curve).  As the case with out-of-chemical-equilibrium thermal distributions, almost the same conclusion can be drawn here. Furthermore, the impacts of the inclusion of scalar sigma states is analyzed. It is obvious that these states seem to enhance the out-of-chemical-equilibrium, especially at large $p_T$.  The excluded-volume corrections are taken into consideration in the way that all stable hadrons and resonances with masses $< 2~$GeV are assumed to equally have finite volume, $r_h$. As discussed earlier, at $r_h=0.2~$fm, the corrections seem very small. There is no such a large variety to increase $r_h$ and simultaneously conserve the thermodynamics consistency. A systematic analysis has been discussed in ref. \cite{Tawfik:2013eua}, see Fig. 1 and the related text. 

The small difference when $r_h$ is increased from $0.2$ to $0.4~$fm can basically be understood from the corresponding freeze-out temperature; with the inclusion of scalar sigma states $T_{ch}\simeq151~$MeV, at $r_h=0.2~$fm. But at $r_h=0.4~$fm, $T_{ch}\simeq167~$MeV. These results are confirmed when scalar sigma states are removed; $T_{ch}\simeq 159~$MeV at $r_h=0.2~$fm. Here, it was not possible to increase $r_h$ to $0.4~$fm due to divergence in the thermodynamic quantities, at the chemical freeze-out, for instance, $T_{ch}$ should jump to $\sim 1~$GeV to fulfill the freeze-out conditions \cite{Tawfik:2004vv}! Therefore we have checked $r_h=0.3~$fm. The corresponding $T_{ch}\simeq 164~$MeV. It seems in order to highlight that these results are limited to the validity of the approximation that all hadrons are conjectured to have the same radius.

We find that at vanishing $\mu_{\pi}$, the results from Eq. (\ref{eq:pionpTa}) with and without scalar sigma states are almost identical. But at finite $\mu_{\pi}$, the HRG calculations with and without scalar sigma states become distinguishable. This can be understood due to the remarkable characteristics and great experimental abundances of the sigma states, which would appear as two-pion scalar-isoscalar resonances and are regarded as excitation of the scalar condensates, i.e. playing similar roles as the Higgs boson does for strong interactions \cite{Shuryak2002}. Their $\bar{q}q$-channel interaction is a maximally attractive one. This could be so strong that it breaks spontaneously the chiral symmetry and produces a quark condensate. Instanton-induced 't Hooft interaction is also conjectured as a mechanism for that attraction \cite{Shuryak2002}.

In order to elaborate more why we have chosen another alternative and thought of confronting measured $p_T$-spectra of pions to Bogolubov superfluidity of Bose-Einstein gas and/or a thermal approach at out-of-chemical equilibrium of pions, i.e. finite $\mu_{\pi}$, some remarks are now in order. Interpreting out-of-chemical-equilibrium in heavy-ion collisions at LHC in terms of the non-extensive Tsallis statistics \cite{jean} should be a subject of a fundamental revision. This has been discussed in great details refs. \cite{Tawfik:2018ahq,Tawfik:2017bsy,Tawfik:2016pwz}. On the one hand, the basic idea behind the thermal approach such as the HRG model is apparently additivity. Obviously, the Tsallis-type approach applies extensive statistics in the {\it additive} HRG, where both exponential and logarithm functions are merely replaced. On the other hand, the $p_T$-spectra of positive pions from the most central ($0-5\%$) Pb+Pb collisions at $2.67~$TeV are not reproducible by Eq. (20) in ref. \cite{jean} (not drawn in Fig. \ref{fig:ptSpect1}). But an excellent reproduction of these $p_T$-spectra is achieved at $\mu_{\pi}=0.12~$GeV, Sec. \ref{sec:thrmlNoNhrgEqlbm}. The excellent agreement apparently covers the entire range of $p_T$, left panel of Fig. \ref{fig:ptSpect1}. 

Middle panel of Fig. \ref{fig:ptSpect1} shows the same in left panel but the $p_T$-spectra are confronted to the HRG model, in which all PDG sigma states are properly included, sec. \ref{sec:hrgEqlbm}. for point-like hadrons and resonances ($r_h=0$), we notice that the agreement is excellent at low $p_T$ and $\mu_{\pi}=0.12~$GeV. At finite volumes ($r_h\neq0$), the reproduction of the $p_T$-spectra of pions within the entire $p_T$-range becomes relatively worst. As seen in left panel, vanishing $\mu_{\pi}$ largely underestimates the $p_T$-spectra, everywhere.

The right panel of Fig. \ref{fig:ptSpect1} draws the same as in middle panel but without PDG sigma states. Almost everywhere except at $p_T<0.2~$GeV, there is an obvious disagreement pointing out to the essential roles that the scalar sigma states seem to play in reproducing $p_T$-spectra of pions at LHC energies.

\section{Conclusions and outlook}
\label{sec:Cncl}

In reproducing the measured $p_T$-spectra of positive pions at LHC energies, we have decided in favor for another alternative thermal approach assuming an out-of-chemical equilibrium of pions, ($\mu_{\pi}\neq0$). In addition to this, we have introduced out-of-chemical equilibrium of pions to the well-known HRG model. In ancillary to the baryon, the strangeness chemical potential and the electric charge potential, we have imposed $\mu_{\pi}\neq0$, as well.

We shortly highlighted the incompleteness of non-extensive Tsallis statistics, especially when confronted to bulk matter created at relativistic energies. The basic idea of implementing an {\it additive} resonance gas even with Tsallis algebra, where both exponential and logarithm functions are properly replaced, seems not at all modifying the extensivity. The latter obviously contradicts the intention of taking into account out-of-chemical equilibrium particle production.

We conclude that $p_T$-spectra of positive pions produced in most central ($0-5\%$) Pb+Pb collisions at $2.67~$TeV are well reproduced at $\mu_{\pi}=0.12~$GeV. This was the case in two different thermal approaches. The first one is based on degeneracy of a non-perfect Bose-Einstein gas or an ingenious application of second quantization known as Bogolubov superfluidity. The second one is the well-known HRG model, in which $\mu_{\pi}\neq0$ was {\it ad hoc} introduced. The baryon, strangeness and electric charge chemical potentials, etc. can also be taken into consideration. This approach seems being successful in reproducing $p_T$-spectra of positive pions, at LHC energies, when PDG sigma states are taken into account.

Future works shall be devoted to investigation of the impacts that the PDG sigma states would have in reproducing $p_T$-spectra of other hadrons. Furthermore, $p_T$-spectra of other bosons shall be extended to understand whether they require out-of-chemical equilibrium processes, as positive pions do. The behavior of pion production at high energies, where the production of kaons, protons and antiprotons seem  requiring anomalously large contributions from the exponential term to describe the shape of their transverse momenta, could be a subject of a similar out-of-chemical-equilibrium analysis.

\section*{Acknowledgment}

The author would like to thank David Blaschke and Ernst-Michael Ilgenfritz for the stimulating discussions on the non-equilibrium pion production at high energies! The author acknowledges the finanial support by DAAD-Widereinladung no. 57378442.

\end{document}